\documentclass[aps,prd,reprint,superscriptaddress,nofootinbib, longbibliography]{revtex4-2}

\usepackage{amssymb}
\usepackage{amsmath}
\usepackage[utf8]{inputenc}
\usepackage{graphicx}
\usepackage{cancel}
\usepackage{bbm}
\usepackage{color}
\usepackage{lineno}
\usepackage{dcolumn}   
\usepackage{bm}        
\usepackage{amssymb}   
\usepackage{amsmath,amssymb}
\usepackage{comment}
\usepackage[dvipsnames]{xcolor}
\usepackage[colorlinks = true,
            linkcolor  = RoyalBlue,
            urlcolor   = RoyalBlue,
            citecolor  = RoyalBlue]{hyperref}
            
\usepackage[utf8]{inputenc}
\usepackage{amsmath}
\usepackage{graphicx}
\usepackage{color}

\begin{document}

\title{A fluid-dynamic approach to heavy-quark diffusion in the quark-gluon plasma}

\author{F. Capellino}
\email[]{f.capellino@gsi.de}
\affiliation{GSI Helmholtzzentrum f{\"u}r Schwerionenforschung, 64291 Darmstadt, Germany}
\affiliation{Physikalisches Institut, Universit{\"a}t Heidelberg, 69120 Heidelberg, Germany}

\author{A. Beraudo}
\email[]{beraudo@to.infn.it}
\affiliation{INFN - Sezione di Torino, via P.Giuria 1, 10125 Torino, Italy}

\author{A. Dubla}
\email[]{a.dubla@cern.ch}
\affiliation{GSI Helmholtzzentrum f{\"u}r Schwerionenforschung, 64291 Darmstadt, Germany}

\author{S. Floerchinger}
\email[]{stefan.floerchinger@uni-jena.de}
\affiliation{Theoretisch-Physikalisches Institut
Friedrich-Schiller-Universität Jena, 07743 Jena, Germany} 

\author{S.~Masciocchi}
\email[]{s.masciocchi@gsi.de}
\affiliation{GSI Helmholtzzentrum f{\"u}r Schwerionenforschung, 64291 Darmstadt, Germany}
\affiliation{Physikalisches Institut, Universit{\"a}t Heidelberg, 69120 Heidelberg, Germany}

\author{J. Pawlowski}
\email[]{j.pawlowski@thphys.uni-heidelberg.de }
\affiliation{Institut f\"{u}r Theoretische Physik, Universit\"{a}t Heidelberg, 69120 Heidelberg, Germany}

\author{I. Selyuzhenkov}
\email[]{ilya.selyuzhenkov@gmail.com}
\affiliation{GSI Helmholtzzentrum f{\"u}r Schwerionenforschung, 64291 Darmstadt, Germany}
\date{\today}

\begin{abstract}
 
 A fluid-dynamic approach to the diffusion of heavy quarks in the quark--gluon plasma (QGP) is presented. Specifically, we analyze the Fokker-Planck equation for the momentum transport of heavy quarks from a fluid perspective and use a mapping to second-order fluid-dynamics to determine conductivities and relaxation times governing their spatial diffusion.
 By investigating the relation between the two approaches, we provide new insights concerning the level of local thermalization of charm and bottom quarks inside the expanding QGP. Our results indicate that a fluid-dynamic description of diffusion is feasible for charm quarks at least for the latest stages of the fireball evolution.
 
\end{abstract}

\maketitle


\section{Introduction}

Relativistic viscous hydrodynamics is a widely used tool in the description of heavy-ion collisions. The assumption of the QGP behaving like an expanding fluid successfully managed to explain light-flavor observables such as particle spectra and flow harmonic coefficients (\cite{GALE_2013,Heinz_2013,Dubla:2018czx}), suggesting that the mean-free path of the light quarks and gluons is substantially smaller than the size of the created fireball. In particular, elliptic flow is an important probe of collectivity in the system created in heavy-ion collisions. It is a response to initial conditions and therefore sensitive to the early and strongly interacting phase of the evolution. Remarkably, recent experimental results~\cite{Acharya_2021,Acharya_2020} show that open heavy-flavor and charmonium states -- D mesons, $J/\psi$ -- have significantly positive elliptic flow. These observations raised questions about  the possible heavy-quark (local) thermalization in the QGP~\cite{Andronic_2021}.

Heavy quarks are very powerful tools to characterize the QGP produced in heavy-ion collisions. Due to their large mass, they are produced via hard scattering processes at the very beginning of the collision and undergo all the stages of the evolution of the expanding medium. The timescale required for the heavy quarks to approach local kinetic equilibrium is expected to be a factor $\sim M/T$ larger than the one for light quarks~\cite{Moore:2004tg}, where $M$ is the heavy-quark mass and $T$ is the temperature of the medium.

A variety of transport models (for a recent review see~\cite{Prino:2016cni}) based on the Boltzmann equation or its approximations were developed in the past years, addressing the issue of heavy-quark in-medium dynamics. These models treat the heavy quarks as Brownian particles that undergo elastic and (possibly) radiative scattering processes with the partons from the QGP. The heavy-quark momentum is modified only slightly during each individual scattering, meaning that several interactions are required in order to change it significantly. This implies that local kinetic equilibrium can be reached by charm and bottom quarks only for rather long timescales. Therefore, recent theoretical developments are mainly focused on the evaluation of the transport coefficients which characterize the medium and parametrize the interaction between the heavy quarks and the light partons from the QGP. Great effort was put towards a consistent estimate of these transport coefficients through a systematic comparison between model predictions and experimental data.

Our purpose here is to address the question of heavy-quark in-medium thermalization from a new point of view. We treat the heavy quarks as part of the medium itself, somehow in analogy to the most recent implementation of the Statistical Hadronization Model \cite{Andronic_2007,Andronic_2021,Andronic:2019wva,Braun_Munzinger_2000}. We assume heavy quarks had enough time to interact with the light thermal partons and to approach local kinetic equilibrium.
This will be found to be a reasonable assumption for charm in the stages in which the expansion of the medium is not too violent.
In this case heavy-quark number conservation laws and the related continuity equations can be used to describe the diffusion dynamics in spacetime.
While one can realistically assume that charm quarks manage to get quite close to local kinetic equilibrium, a chemical thermalization would only happen over larger time scales: hence the heavy quark multiplicity is set by the initial production in hard scattering processes and remains almost unchanged during the medium evolution.

In spite of introducing new dissipative quantities to initialize, such a theoretical description is still much more economic than a numerical solution of the Boltzmann equation. 

Eventually, heavy quarks could even affect the dynamics of the QGP itself and this could be naturally encoded into a system of coupled hydrodynamic equations. However, as a starting point, it is reasonable to think that the heavy quarks do not influence significantly the energy density, pressure, velocity or shear stress of the medium. These quantities are mostly determined by the thermodynamics of the light quarks and gluons degrees of freedom. The heavy quarks can be added "on top" of the fluid and their fluid dynamics is described by additional conserved currents. In the following we pursue such an "on top" description.
By studying the connection between hydrodynamics and transport theory (Fokker-Planck equation) we obtain new insights on the mechanisms of ``hydrodynamization'' of the heavy degrees of freedom in the QGP medium.

This work is structured as follows.
In Sec.~\ref{sec:FokkerPlanck} we present the aspects of transport theory which are relevant for our treatment and give a short  overview about heavy-quark transport coefficients in the literature. In Sec.~\ref{sec:hydro} we introduce the conserved currents associated to the heavy quark propagation in the QGP. In Sec.~\ref{sec:merging} we build a relation between the transport coefficients defined in the hydrodynamic framework and the ones in transport theory. In Sec.~\ref{sec:results} we show the results for the numerical evaluation of the hydrodynamic transport coefficients. In Sec.~\ref{sec:Bjorken} we test the validity of the hydrodynamic description of heavy quarks in the case of a QGP undergoing Bjorken flow. 
Finally, in Sec.~\ref{sec:conclusions} we draw our conclusions and discuss possible developments and future perspectives.

\section{\label{sec:FokkerPlanck}The Fokker-Planck equation and heavy-quark transport coefficients}

In this section we present the Fokker-Planck equation as an approximation of the Boltzmann equation and we give an overview about heavy-quark transport coefficients.

The Boltzmann equation relates the change in time of the (out-of-equilibrium) distribution function $f_k$ of a certain particle, with momentum $k$, to the collision integral $C[f_k]$,
\begin{equation}
    k^\mu\partial_\mu f_k = C[f_k]
.\end{equation}
Let us consider the collision integral for the elastic scattering between a heavy quark of initial momentum $k$ and a light parton from the medium of initial momentum $k'$. Denoting the outgoing momenta of the heavy quark and parton with $p$ and $p'$, the collision integral in its classical form then reads
\begin{equation} 
    C[f^{(r)}_{k}]=\int dK'dPdP'\,W_{kk'\rightarrow pp'}\Bigl(f^{(r)}_pf_{p'}-f^{(r)}_{k} f_{k'}\Bigr)
\,,\end{equation} 
where $W_{kk'\rightarrow pp'}$ is the scattering rate for the aforementioned process and $r$ is an index accounting for the heavy quark (charm/bottom) or antiquark (anticharm/antibottom). 
We employed the abbreviation
\begin{equation}
    \int dP = g \int \frac{d^3 p}{(2\pi)^3 p^0} 
\,,\end{equation}
to indicate the integral over the phase space of a particle with four-momentum $p^\mu$. The degeneracy factor $g$ accounts for internal degrees of freedom (spin, color etc) and the time component of the four-momentum $p^0$ is evaluated on-shell.

The Fokker-Planck equation is an approximation of the Boltzmann equation in the limit of \textit{multiple soft scatterings} between the heavy quark and a parton from the medium. We recall their relation in a situation where the fluid of gluons and light quarks is stationary and homogeneous. This is sufficient to define the transport coefficient of interest for this work. The collision integral expanded in terms of the small transferred momentum $q=p-k$ up to second order in momentum derivatives reads
\begin{equation}
\label{eqn:fokker-planck}
     C[f^{(r)}_{k}] = k^0\frac{\partial}{\partial k^i}\left\{
     {A^i}f^{(r)}_{k}+\frac{\partial}{\partial k^j}\left[{B^{ij}}f^{(r)}_{k}\right]
     \right\}
\,,\end{equation}
where the indices $i,j=1,2,3$ run over the spatial components of the correspondent four-momentum vector. The tensors $A^i$ and $B^{ij}$ describing the interaction of the heavy quark with the medium arise naturally from the momentum expansion of the collision integral.
For an isotropic medium, after factorizing the tensorial structure, they can be rewritten as
\begin{equation}
\begin{aligned}
    A^i(\vec{k}) =&\, A({k})k^i\,,\\[1ex]
    {B^{ij}(\vec{k})} =&\, (\delta^{ij}-\hat{k}^i\hat{k}^j){B_0(k)} + \hat{k}^i\hat{k}^j{B_1(k)}
\,,\end{aligned}
\end{equation}
where we used $\vec{k}$ to indicate the spatial part of the four-momentum vector $k^\mu$ and $\hat{k}^i \equiv k^i/|\vec{k}|$, with $k\equiv|\vec{k}|$. $A(k)$ represents a \textit{drag} coefficient and $B_0$ and $B_1$ play the role of \textit{momentum-diffusion} coefficients along the directions orthogonal and parallel to the heavy-quark velocity, respectively (for a full derivation see e.g. Ref.~\cite{Svetitsky:1987gq}). The Einstein fluctuation-dissipation (EFD) relation for the three transport coefficients reads
\begin{equation}
    A=\frac{1}{T k^0} B_1 -\frac{1}{k^2}\left[2(B_1-B_0)+k\frac{\partial B_1}{\partial k} \right]
,\end{equation}
in three spatial dimensions, where $T$ is the temperature of the surrounding medium and $k^0\equiv\sqrt{k^2+M^2}$ is the energy of the on-shell heavy quark~\cite{Reichl,Kubo}. This ensures that, asymptotically, the heavy quark momentum distribution approaches the Maxwell-J\"uttner limit $f_k^{(r)}\sim e^{-k^0/T}$. An extension accounting for quantum corrections for the heavy quarks (Pauli blocking) asymptotically approaching a Fermi-Dirac distribution is discussed in Appendix \ref{app:quantumFP}.

Very often in phenomenological studies one attempts to summarize the heavy-quark coupling with the medium in terms of a single coefficient, the spatial diffusion coefficient $D_s$, identified via the asymptotic mean squared displacement $\langle\vec x^2\rangle\underset{t\to \infty}{\sim} 6 D_s t$ of an ensemble of heavy quarks initially placed at the origin. One can show that, as long as the dynamics is non-relativistic ($M\gg T$), the latter is related to the other transport coefficient by
\begin{gather}
\label{def:Ds}
    D_s=\lim_{k\rightarrow 0}{\frac{T}{M A(k)}}\,
    ,\end{gather}
where $k$ is the heavy-quark momentum and $M$ is its mass. 
Recent constraints $1.5 < 2\pi D_sT_{pc} < 4.5 $ at the pseudocritical temperature $T_{pc}$ = 0.155 GeV \cite{ALICE:2021rxa} were obtained by fitting various transport models to ALICE experimental data for the nuclear modification factor $R_{AA}$, elliptic ($v_2$) and triangular ($v_3$) flow of D mesons in Pb-Pb collisions at $\sqrt{s_{NN}}=5.02$ TeV. This constraint corresponds to a thermalization time of about $\sim 3-9$ fm/c at $T_{pc}$ = 0.155 GeV for charm quarks of mass $M=1.5$ GeV (the link between $D_s$ and the relaxation time will be better clarified in the following).
In this work, we assume that this estimate for $D_s$ is also applicable at temperatures above $T_{pc}$ and for bottom quarks.
Beside this phenomenological estimate, in this work we also employ lattice-QCD (lQCD) results for $D_s$ taken from Ref.~\cite{Altenkort:2021umr}. Other results can be found in Refs.~\cite{Altenkort:2020fgs,Francis-2015-lQCD,Ding:2021ise}.
The lattice-QCD results for $D_s$ (so far limited to the quenched approximation) used in this paper arise from calculations of color-electric field correlators performed in the static $M\to\infty$ limit, which provide the momentum broadening of an infinitely heavy quark
(for more details see~\cite{Rapp:2018qla} and references therein). 
The above estimates for $D_s$ are expected to be more reliable for objects with a larger mass, such as bottom quarks. Nevertheless, as done in Ref.~\cite{Francis-2015-lQCD}, one can attempt to apply these results to charm quarks as well. Notice that the quenched approximation requires a global adaptation of scales from pure Yang-Mills theory to QCD. Such a procedure was suggested in~\cite{Christiansen:2014ypa} for the shear viscosity over entropy ratio $\eta/s$, and will be applied to the heavy-quark transport coefficients in a continuation of the present work.

An alternative approach towards the computation of heavy-quark transport coefficients has been suggested recently, based on the theory of open quantum systems~\cite{Brambilla:2019tpt} within the EFT framework of potential non-relativistic QCD (pNRQCD) calculations~\cite{Brambilla:1999xf}. It avoids some of the above limitations, and in particular it allows for lattice estimates of heavy-quark transport coefficients beyond the quenched approximation.

\section{The hydrodynamic approach to heavy quarks}
\label{sec:hydro}
The aim of the present section is to introduce the heavy-quark conserved current which propagates causally in the QGP.
As discussed in the previous section, the transport coefficients fitted to reproduce experimental data or estimated from lattice-QCD simulations suggest the possibility at least for charm quarks to approach kinetic equilibrium during their propagation in the hot fireball arising from the nuclear collision. 
Hence it looks reasonable to develop a hydrodynamic approach also for the modeling of the heavy-quark dynamics in the quark-gluon plasma, that we are going to present in this section.

It is crucial to construct the hydrodynamic approach such that causality is preserved even in the presence of dissipative effects associated to the finite mean-free-path of the plasma particles. We employ an Israel-Stewart type formalism -- or \textit{second-order} hydrodynamics -- in which the dissipative quantities (the heavy-quark diffusion currents in this case) are promoted to dynamical variables which evolve according to certain equations of motion. Here these equations are governed by conductivities and relaxation times. The relaxation times have to be large enough in order to prevent the non-causal behavior; at the same time they have to be smaller than the inverse expansion rate of the fireball (coinciding with the longitudinal proper time $\tau$ for a pure longitudinal Bjorken expansion) in order for the hydrodynamic approach to hold. What the relaxation time tells us is that we are dealing with out-of-equilibrium transient hydrodynamics for a time scale of the order of the relaxation time itself. This relaxation towards a hydrodynamic phase is often called \textit{hydrodynamization} (see e.g. \cite{Heller:2016rtz}). 

In our specific problem, we want to include the conservation of a heavy quark--antiquark ($Q\bar Q$) current. 
Two relevant heavy-quark currents are
\begin{equation}
    N^\mu_+\equiv\frac{N^\mu_Q+ N^\mu_{\bar Q}}{2}\quad {\rm and}\quad N^\mu_-\equiv N^\mu_Q- N^\mu_{\bar Q}\,, \label{eq:QQbardef}
\end{equation}
associated to the conservation of the average (+) and net (-) heavy-quark number, respectively.
Notice that in the situation of experimental interest the net heavy-quark number vanishes and their average number coincides with the number of $Q\bar Q$ pairs initially produced in the hard scattering processes and conserved throughout the fireball evolution.
The number of $Q\bar Q$ pairs is expected to be \textit{accidentally} conserved during the evolution of the QGP. The mass of the heavy quarks is too large for them to be thermally produced~\cite{Braun-Munzinger:2007fth}.
At the same time their annihilation rate is too small to lead to a measurable loss of $Q\overline Q$ pairs during the short lifetime of the plasma. Hence their final multiplicity is fixed by the initial production in hard partonic processes described by pQCD.
On the other hand, the net heavy-quark number is expected to be \textit{exactly} conserved in QCD due to the symmetry of the interaction. The loss of a single quark/antiquark is in fact forbidden by flavor conservation. The net heavy-quark current is not conserved by electroweak interactions instead. However, electroweak processes can be considered negligible  within the lifetime of the QGP since they require much longer timescales.

Since the numbers of heavy quarks and antiquarks are separately conserved  within the fireball lifetime, following the work in
Ref.~\cite{Denicol_2012}, we write the corresponding conserved currents including dissipative corrections as
\begin{equation}
\begin{aligned}
    & N^\mu_{(r)} = n_{(r)} u^\mu + \nu^\mu_{(r)}\,,\\
    & \partial_\mu N^\mu_{(r)} = 0\,.
\end{aligned}
\end{equation}
Here $r\!=\!Q$ or $\overline Q$, $u^\mu$ is the fluid four-velocity and $\nu^\mu_{(r)}$ are the heavy-(anti)quark diffusion currents, constructed to be orthogonal to $u^\mu$, i.e. $u_\mu\nu_{(r)}^\mu=0$. Notice that this last condition entails that in the LRF of the fluid -- in which $u^\mu = (1,0,0,0)$ -- the time component of the diffusion currents $\nu_{(r)}^0$ vanishes. In this frame the time component of the current $N^\mu_{(r)}$ defines then the heavy-(anti-)quark density $n_{(r)}$ even in the presence of dissipative corrections.
At local kinetic equilibrium, we consider for quarks and antiquarks a Boltzmann distribution,
\begin{equation}
\begin{split}
\label{eqn:boltzmanndistribution}
    f^{(r)}_{0k}\!=\!\exp{\left(\frac{-E_k +\mu_r}{T}\right)}\!\\
    =\!\exp{\left(\frac{-E_k \!+\!q_r\mu^{\rm net}_{Q} \!+\! \mu_Q^{\rm ave}/2}{T}\right)}
,
\end{split}
\end{equation}
where $E_k=u_\mu k^\mu$.  
The $\mu^{\rm net}_{Q}$ is the chemical potential associated to net $N^\mu_-$ conserved current and $q_r$ is a charge factor -- positive for quarks and negative for antiquarks. 
Additionally, one should consider that heavy quarks are produced out of chemical equilibrium in the QGP and their number is conserved during the subsequent evolution of the fireball.
A chemical potential $\mu_Q^{\rm ave}$, the same for quarks and antiquarks,  associated to their average number must be included in order to account for such a deviation from full thermodynamic equilibrium. In summary, one has
\begin{eqnarray}
\mu_Q&=&\mu_Q^{\rm ave}/2+\mu_Q^{\rm net}\,,\nonumber\\
\mu_{\overline Q}&=&\mu_Q^{\rm ave}/2-\mu_Q^{\rm net}\;,
\end{eqnarray}
consistently with the thermodynamic identities
\begin{equation}
    n_r=\frac{\partial P}{\partial\mu_r},\quad n_-=\frac{\partial P}{\partial\mu_Q^{\rm net}},\quad
    n_+=\frac{\partial P}{\partial\mu_Q^{\rm ave}}\;.
\end{equation}

It is often convenient to introduce the heavy-quark fugacity $\gamma_Q\equiv e^{\mu_Q^{\rm ave}/2T}$ which can be factored out from the heavy (anti)quark distributions:
\begin{equation}
    f^{(r)}_{0k}\!=\!\gamma_Q\exp{\left(\frac{-E_k \!+\!q_r\mu^{\rm net}_{Q}}{T}\right)}\;.
\end{equation}

In Appendix \ref{app:fugacityStudy} we provide an estimate of $\gamma_Q$ in the case of a fluid undergoing Bjorken flow. 
In the following, we simply focus on the conservation of the average heavy-quark number, since in most cases one is not interested in distinguishing hadrons arising from a $Q$ or $\overline Q$ parent parton (an exception could be the difference $\Delta v_1$ in the direct flow of $D^0$ and $\overline D^0$ mesons proposed as a tool to extract information on the primordial magnetic field in the plasma~\cite{Das:2016cwd}).
Furthermore, for simplicity we assume that $N^\mu_-=0$, i.e. $\mu_Q^{\rm net}=0$, since the initial hard processes lead to the production of the same number of quarks and antiquarks and we neglect any local unbalance developing during the hydrodynamic evolution.
We define then $\sum_r n_0^{(r)}/2\equiv n_+$ and $\sum_r \nu^\mu _{(r)}/2\equiv \nu^\mu_+$. In this case, the dynamic evolution of the {  relevant} diffusion current will be driven by a single chemical potential $\mu_Q=\mu_{\overline Q}=\mu_Q^{\rm ave}/2$. 
We look for an equation of motion for the particle diffusion current in the form
\begin{equation}
\label{eqn:eom_partdiffcurr}
     \tau_n \Delta^\mu_{\,\rho} u^\sigma \partial_\sigma  \nu_+^\rho + \nu^{\mu}_+= \kappa_n  \nabla^\mu \left( \frac{\mu_Q}{T} \right)\,,
\end{equation}
where $\Delta^{\mu \nu}=g^{\mu \nu}-u^\mu u^\nu$ is the projector onto the space orthogonal to the fluid velocity and we defined the transverse gradient $\nabla^\mu\equiv\Delta^{\mu\nu}\partial_\nu$.
This is a relaxation-type equation in which terms of higher order in the gradients are neglected.
Two transport coefficients are present in Eq.~\eqref{eqn:eom_partdiffcurr}, namely the relaxation time $\tau_n$ and the particle-diffusion coefficient $\kappa_n$. The presence of a relaxation time, as anticipated, is necessary in order to ensure the causality of the equation. 
For $\tau\gg\tau_n$, $\nu^\mu_+$ relaxes to its Navier-Stokes limit
$\nu^\mu_+=\kappa_n \nabla^\mu \left( {\mu_Q}/{T}\right)$.

\section{Heavy-quark relaxation time and transport coefficients}
The purpose of this section is twofold. First we study the relation between the transport coefficients defined in the hydrodynamic approach and the ones defined in transport theory (Fokker-Planck equation). Secondly, we show our numerical results for the hydrodynamic transport coefficients.

\subsection{Matching Fokker-Planck with hydrodynamics}
\label{sec:merging}
The definition of the heavy-quark relaxation time $\tau_n$ and diffusion coefficient $\kappa_n$ are deeply related to the collision integral entering the Boltzmann equation. One can start from the Fokker-Planck equation for the heavy (anti)quark distribution $f_k^{(r)}$, written for the case of a homogeneous fluid at rest, and integrate subsequent moments of it, taking at the end the proper linear combination to get an equation for the diffusion current $\nu_+^\mu$. The zeroth moment simply gives the conservation or
continuity equation, which, in the fluid rest frame, reduces to
\begin{equation}
\label{eqn:continuity}
   \partial_t n_+ +\partial_i \nu^i_+ = 0\,.
\end{equation}

The first moment gives
\begin{align}\nonumber 
&\partial_t \int dK k^0 k^l f_k^{(r)} + \partial_i\int dK {k^l k^i} f_k^{(r)} \\[1ex]
=&\int dK k^l\left(k^0\frac{\partial}{\partial k^i}\left\{
     {A^i}f^{(r)}_{k}+\frac{\partial}{\partial k^j}\left[{B^{ij}}f^{(r)}_{k}\right]
     \right\}\right)    \,.
\end{align}

We use the following decomposition for $f_k^{(r)}$,
\begin{equation}
    f_k^{(r)} = f_{0k}^{(r)} + \delta f_k^{(r)}\,.
\end{equation}
where in the equilibrium part we allow the chemical potential to depend on the spacetime point $x$, allowing for the development of a local excess of heavy quarks.
In the following we employ a simplified version of the approach developed by Denicol \emph{et al.} in Ref.~\cite{Denicol_2012}. Hence, the deviation from local equilibrium $\delta f_k^{(r)}$ is expanded in terms of its moments,
\begin{equation}
     \rho^{\langle \mu_1 ..\mu_l \rangle}_{(r)} \equiv \Delta^{ \mu_1 ..\mu_l}_{\, \nu_1 ..\nu_l} \int dK k^{\langle \nu_1}..k^{\nu_l \rangle} \delta f_k^{(r)}\,,
\end{equation}
as follows
\begin{equation}
\label{eq:deviation_exp}
    \delta f_k^{(r)} = f_{0k}^{(r)}\left(\sum_{l=0}^{\infty} a_{l}^{(r)} \rho^{\mu_1 ..\mu_l}_{(r)} k_{\langle \mu_1}..k_{\mu_l \rangle}\right)\,,
\end{equation}
where $a_l^{(r)}$ are the coefficients of the linear expansion.  The projectors $\Delta^{ \mu_1 ..\mu_l}_{\, \nu_1 ..\nu_l}$ to the fully symmetric, transverse and traceless part of a tensor are defined as in~\cite{DeGroot:1980dk,Denicol_2012}. 
Given a tensor $A^{\nu_1 ..\nu_l}$, by applying the projector $\Delta^{ \mu_1 ..\mu_l}_{\, \nu_1 ..\nu_l}$ one obtains
\begin{equation}
    \label{eqn:projector}
    A^{ \langle \mu_1 ..\mu_l \rangle}\equiv\Delta^{ \mu_1 ..\mu_l}_{\, \nu_1 ..\nu_l} A^{\nu_1 ..\nu_l}\,.
\end{equation}
{Stopping the expansion at second order one only needs the usual transverse projector $\Delta^\mu_\nu$ and
\begin{equation}
    \Delta^{\mu_1\mu_2}_{\nu_1\nu_2}\equiv\frac{1}{2}(\Delta^{\mu_1}_{\nu_1}\Delta^{\mu_2}_{\nu_2}+\Delta^{\mu_1}_{\nu_2}\Delta^{\mu_2}_{\nu_1})-\frac{1}{3}\Delta^{\mu_1\mu_2}\Delta_{\nu_1\nu_2}\;.
\end{equation}}
According to this definition, one has
\begin{equation}
\begin{aligned}
    \rho_{(r)} &= -\frac{3}{M^2} \Pi_{(r)}\,,\\
    \rho^\mu_{(r)} &= \nu^\mu_{(r)}\,,\\
    \rho^{\mu \nu}_{(r)} &= \pi^{\mu \nu}_{(r)}\,,
\end{aligned}   
\end{equation}
being respectively the bulk pressure, the diffusion current and the shear stress tensor associated to the heavy (anti)quarks.
In getting these results one has exploited the Landau matching conditions, which ensure that
\begin{equation}
    \int dK(k\!\cdot\! u)\delta f_k^{(r)}=0\quad{\rm and}\quad
    \int dK(k\!\cdot\! u)^2\delta f_k^{(r)}=0\,.
\end{equation}
They are a way of fixing a temperature and chemical potential of the system, even when the latter is off-equilibrium, starting from the knowledge of the particle and energy density, obtained from the first two moments of the particle distribution.

By neglecting moments $\rho^{\mu_1..\mu_l}_{(r)}$ of rank higher than 2, the dissipative correction to the heavy-quark distribution reads then
\begin{equation}
\begin{split}
\delta f_k^{(r)} = f_{0k}^{(r)} \Big(-a_0^{(r)}\frac{3}{M^2} \Pi_{(r)} + a_1^{(r)}\nu^\mu_{(r)}\, k_{\langle \mu \rangle}  + \\+
a_2^{(r)} \pi^{\mu \sigma}_{(r)}\, k_{\langle \mu} k_{\sigma \rangle}\Big)\,.
\end{split}
\end{equation}
In the expression above one can determine the coefficients $a_l^{(r)}$ exploiting the definition of the bulk pressure, diffusion current and shear stress in terms of the first three moments of $\delta f_k^{(r)}$ respectively (see Appendix \ref{app:calculations}) obtaining
\begin{equation}
    a_0^{(r)} = \frac{1}{I_{00}^{(r)}}\,,\quad
    a_1^{(r)} = -\frac{1}{P_0^{(r)}}\,,\quad
    a_2^{(r)} = \frac{1}{2 I_{42}^{(r)}}\,,
\end{equation}

where $P_0^{(r)}$ is the heavy-quark contribution to the pressure and the thermodynamic integrals $I_{nq}^{(r)}$, for the case of a medium at rest, are defined according to Ref.~\cite{Denicol_2012} as 
\begin{equation}
    I_{nq}^{(r)}= \frac{1}{(2q+1)!!} \int dK (k^0 )^{n-2q}k^{2q} f_{0k}^{(r)}\,.
\end{equation}
Notice that the bulk pressure and shear-stress associated to the heavy (anti)quarks are expected to be much smaller than the ones appearing in the stress-energy tensor of the fluid dominated by gluons and light quarks. Furthermore they will enter in the equation for the heavy-quark diffusion current only through their derivatives, providing corrections at least of second order in the gradients. Thus, we will neglect them in our treatment. 
We can then approximate
\begin{equation}
\delta f_k^{(r)} \approx -\frac{1}{P_0^{(r)}}f_{0k}^{(r)} \nu^\mu_{(r)}\, k_{\langle \mu \rangle}\,.
\end{equation}
At first order in the gradients (i.e. neglecting bulk and shear corrections -- see Appendix \ref{app:bulk_corrections} for more details on the calculations) we find a relaxation-type equation for the diffusion current of the form of Eq.~\eqref{eqn:eom_partdiffcurr},  where the transport coefficients read
\begin{align}
     \tau_n = &\, \frac{I_{31}^{(r)}}{(1/3)\int dK k^0 A\, k^2 f_{0k} ^{(r)}}\,,\\
      \kappa_n = &\, \frac{P_0^{(r)} T}{(1/3)\int dK k^0 A k^2 f_0^{(r)}}\,n_0^{(r)}\,.
\end{align}
If one neglects the momentum dependence of the momentum-diffusion coefficients, assuming $D\!\equiv\! B_0\!=\!B_1$, which is shown to be a reliable approximation up to heavy-quark momentum $k \sim$ 5 GeV for bottom quarks~\cite{Alberico_2011},  and imposing the Einstein relation $A(k)=D/E_kT$ one obtains
\begin{gather}
\label{eqn:taun}
    \tau_n = \frac{T I_{31}}{D P_0} = \frac{D_s  I_{31}}{T P_0}\,,\\
    \kappa_n = \frac{T^2}{D} n_0^{(r)} = D_s n_0^{(r)}  \,.
\end{gather}
where we find that the relation $D_s = T^2/D$ between the spatial ($D_s$) and momentum ($D$) diffusion coefficients, usually found in studying the non-relativistic Brownian motion, arises naturally and holds also in this case in which the heavy particle undergoes a relativistic dynamics, with $E_k=\sqrt{k^2+M^2}$. This is a non-trivial result, valid as long as the momentum dependence of $D$ can be neglected.
The index $r$ in Eq.~\eqref{eqn:taun} was omitted since the ratio $I_{31}/P_0$ is equal for quarks and antiquarks.
Notice that in the non-relativistic limit we have
\begin{equation}
\begin{aligned}
    k^0 \sim&\,M\,,\\[1ex]
    I_{31} \sim&\, M P_0\,,
\end{aligned}
\end{equation}
and thus $\tau_n = A^{-1}$. This represents an important consistency check, since $\tau_n$ approaches, in the $M\gg T$ limit, the well known result for the relaxation time arising from the solution of the non-relativistic Fokker-Planck equation.

\subsection{Heavy-quark relaxation time}
\label{sec:results}

\begin{figure}[ht!]
    \centering
    \includegraphics[width=0.5\textwidth]{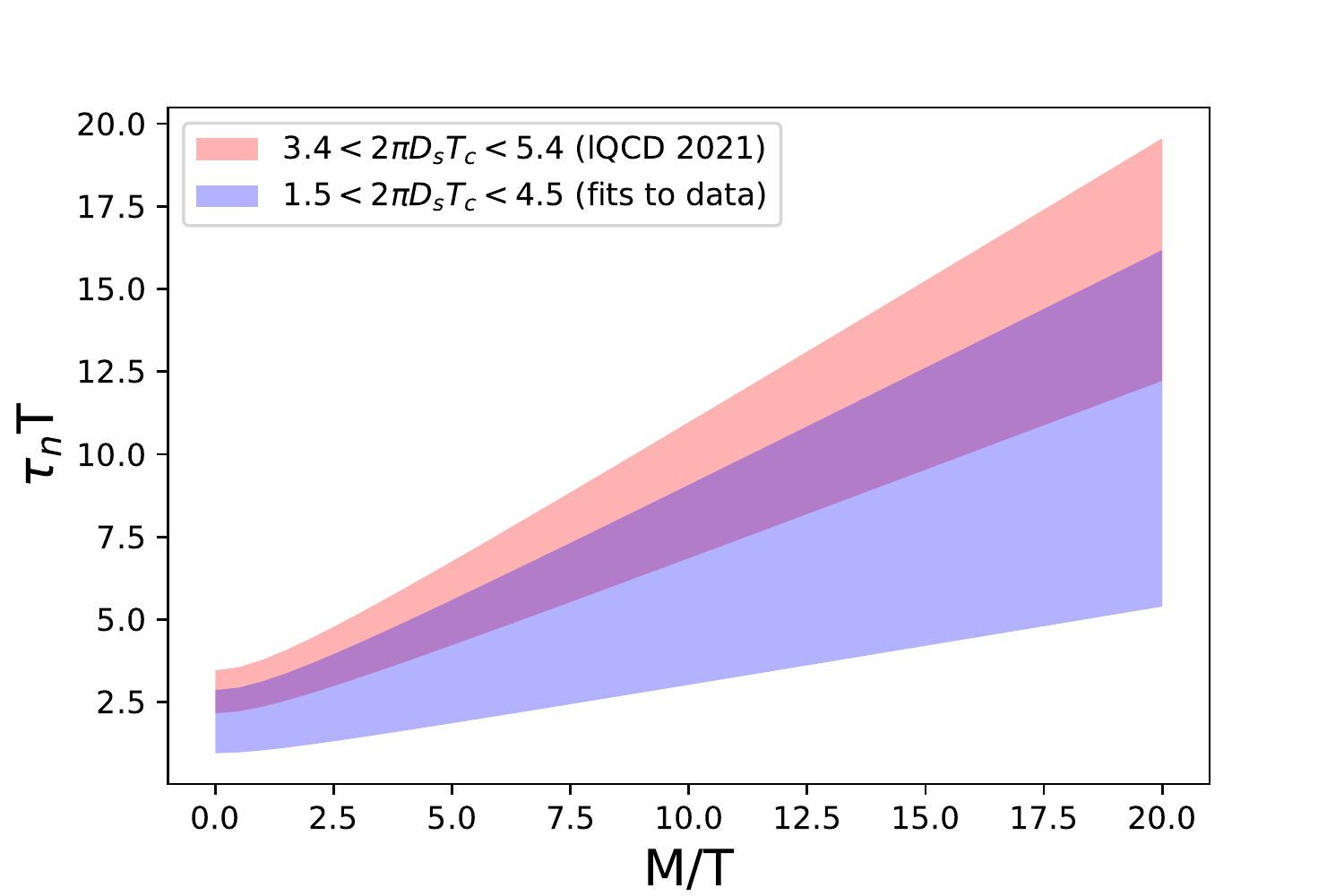}
    \caption{Heavy-quark relaxation time $\tau_n$ multiplied by temperature $T$ as a function of $M/T$. The red band is computed using $D_s$ estimates coming from Lattice-QCD simulations~\cite{Altenkort:2021umr}. The blue band is computed using estimates for $D_s$ coming from fits to ALICE experimental data~\cite{ALICE:2021rxa}.}
    \label{fig:tau_MT}
\end{figure}

In Fig.~\ref{fig:tau_MT} the relaxation time $\tau_n$ multiplied by the temperature is shown as a function of the ratio $M/T$. Here and in the next plots the range spanned by $\tau_n$ is highlighted by the colored bands. Different colored bands correspond to different $D_s$ estimates coming from lattice-QCD simulations~\cite{Altenkort:2021umr} and from fits to ALICE experimental data ~\cite{ALICE:2021rxa}. 
The heavy-quark relaxation time increases linearly with the $M/T$ ratio when the latter is large enough. At a given temperature, the relaxation time is then larger for heavier quarks, as expected. This entails that the non-hydrodynamic phase is lasting longer for bottom quarks with respect to charm quarks. 
The relaxation time $\tau_n$ is observed to be positive even at zero mass. This observation, although referring to a limiting case outside the domain of validity of our approximations, is in agreement with the second-order hydrodynamic description and guarantees causal propagation. 

\begin{figure}
    \centering
    \includegraphics[width=0.5\textwidth]{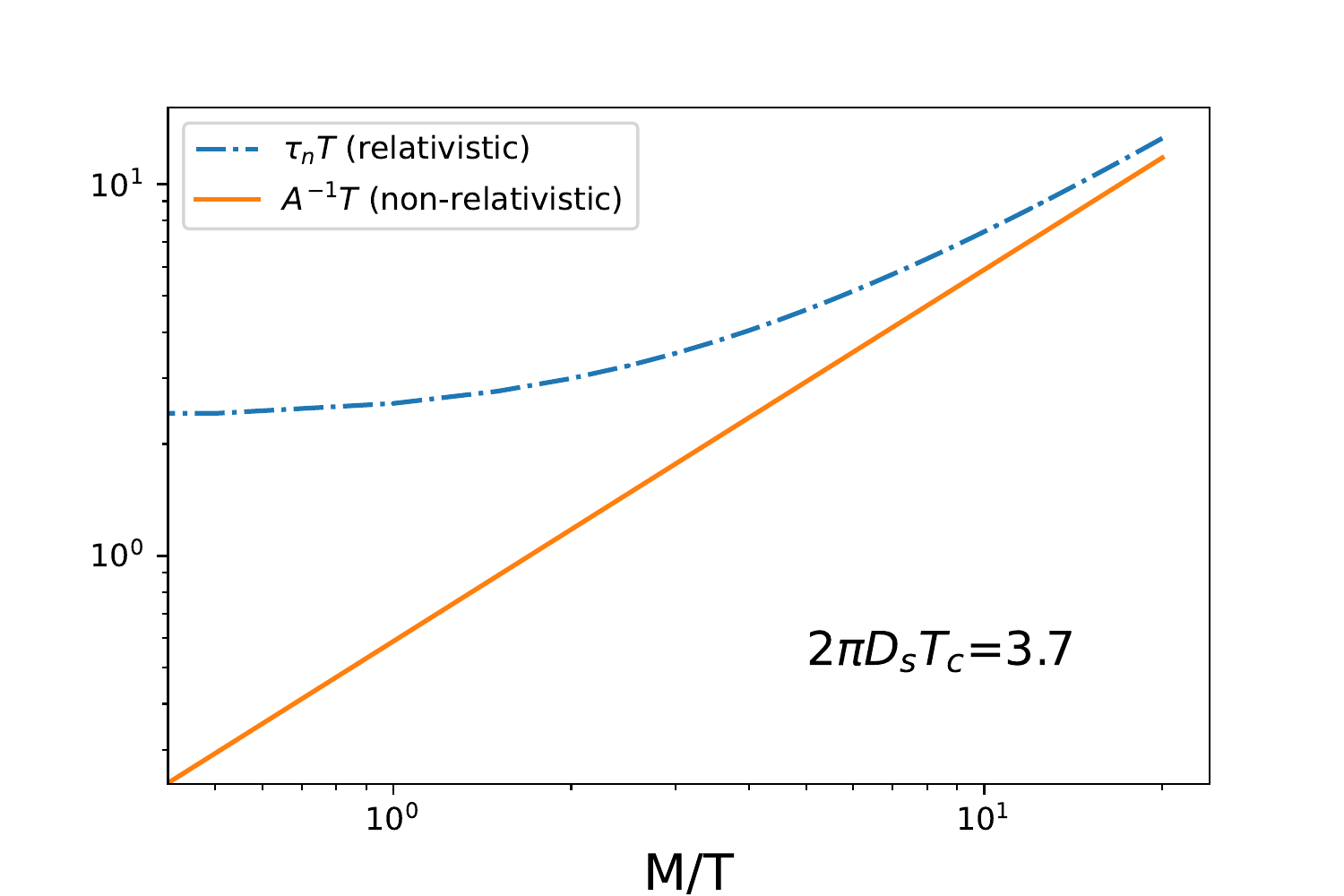}
    \caption{The heavy-quark relaxation time as a function of $M/T$ is compared to the inverse of the Fokker-Planck drag coefficient $A$. Dimensionless units rescaled by the temperature are employed. The two curves coincide in the non-relativistic limit, i.e. for large values of $M/T$.}
    \label{fig:nrlimit}
\end{figure}

In Fig.~\ref{fig:nrlimit} we compare -- in dimensionless units rescaled by the temperature -- our estimate for the relaxation time $\tau_n$ with the inverse of the Fokker-Planck drag coefficient $A$ {arising from the non-relativistic Einstein fluctuation-dissipation relation $A^{-1}\!=\!(M/T) D_s$}. They are both computed according to a spatial diffusion coefficient given by $2 \pi D_s T_c = 3.7$, which falls in both the Lattice-QCD and ALICE ranges. If one assumes  that this last estimate holds also at higher temperatures, the results plotted in Fig.~\ref{fig:nrlimit} do not depend on the specific value of $T$. Notice that one can recast the non-relativistic Einstein relation in a dimensionless form suited to highlight its linear $(M/T)$ scaling
\begin{equation}
    A^{-1}T=\frac{1}{2\pi}\left(\frac{M}{T}\right)(2\pi D_s T)\,,\label{eq:Einstein-rescaled}
\end{equation}
manifest in Fig.~\ref{fig:nrlimit}.
We observe that for large values of $M/T$ the two curves coincides, hence our calculation for the heavy-quark relaxation time $\tau_n$ leads to the correct non-relativistic limit, allowing one to get at the same time a more realistic estimate for the latter in a kinematic range in which the non-relativistic approximation is no longer justified.

\section{Validity of the hydrodynamic description of heavy quarks}
In this section we test the validity of the fluid-dynamic description of heavy quarks in the case of a QGP undergoing Bjorken flow.

\label{sec:Bjorken}
In order to estimate whether it is conceivable for the heavy quarks to be described by fluid-dynamics within an expanding medium before the freeze-out occurs, the relaxation time $\tau_n$ of charm and bottom quarks is compared with the typical expansion time $\tau_{\mathrm{exp}}$ of the fluid, defined as the inverse of its expansion rate. One would be able to treat heavy-quark transport with hydrodynamics only if $\tau_n \ll \tau_{\mathrm{exp}}$. We assume here the fluid expansion to be described by the Bjorken flow model~\cite{Bjorken1983}, in which a purely longitudinal expansion along the beam axis is considered. The system is assumed to be invariant under longitudinal Lorentz boosts and the velocity profile has the form of a Hubble-law expansion along the beam axis $z$,
\begin{equation}
	v_x = v_y = 0\,, \quad v_z = \frac{z}{t}.
\end{equation}
Moreover, in the Bjorken framework, all the thermodynamic quantities depend only on {$\tau\equiv\sqrt{t^2-z^2}$}, that is, the \textit{longitudinal proper time} measured by a clock in the local rest frame of the fluid. {In the case of an ideal expansion, due to entropy conservation}, the temperature follows the power law
\begin{equation}
  T(\tau) = T_0 \left(\frac{\tau_0}{\tau}\right) ^{\frac{1}{3}}  
,\end{equation}
with $T_0 = T(\tau_0)$ being the temperature of the system at $\tau_0$ (formation time of the QGP). The expansion rate of the fluid in the case of this simple flow is given by $\theta=\nabla_\mu u^\mu=1/\tau$, so the typical expansion time-scale is $\tau_{\mathrm{exp}}\equiv 1/\theta=\tau$, coinciding with the longitudinal proper time.
Before displaying our numerical results we can attempt some parametric estimates for the heavy-quark relaxation time arising from the Einstein Fluctuation-Dissipation relation in Eq.~(\ref{eq:Einstein-rescaled}) under the assumption that the product $D_sT$ remains constant. One has
\begin{equation}
    \tau_Q^{\rm EFD}\equiv A^{-1}\sim 1/T^2\sim\frac{1}{(T_0^3\tau_0)^{2/3}}\,\tau^{2/3}\,.\label{eq:tauQ_FDR}
\end{equation}
Hence, for large enough time, one has
\begin{equation*}
\tau_Q^{\rm EFD}\sim\tau^{2/3}<\tau_{\rm exp}=\tau\;.
\end{equation*}
If this occurs before hadronization, at least for a fraction of the fireball lifetime the heavy-quark evolution can be described by hydrodynamic equations, as the other conserved quantities.

\begin{figure}[bht!]
    \centering
    \includegraphics[width=0.5\textwidth]{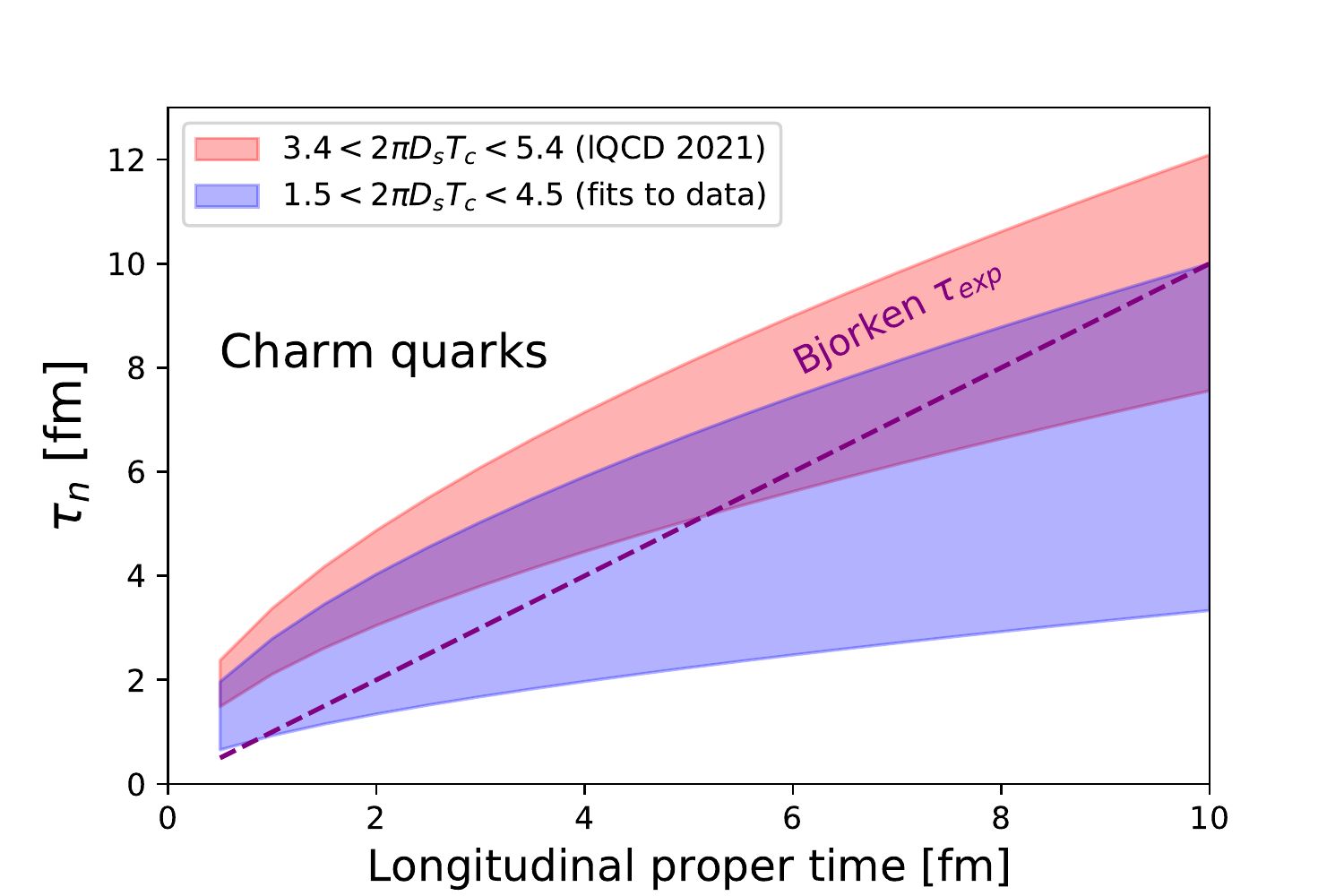}
    \caption{The relaxation time $\tau_n$ of charm quarks as a function of the longitudinal proper time is compared to the typical expansion timescale $\tau_{\rm exp}$ of the fluid undergoing a Bjorken flow. 
    }
    \label{fig:charm_bjorken}
\end{figure}

\begin{figure}[bht!]
    \centering
    \includegraphics[width=0.5\textwidth]{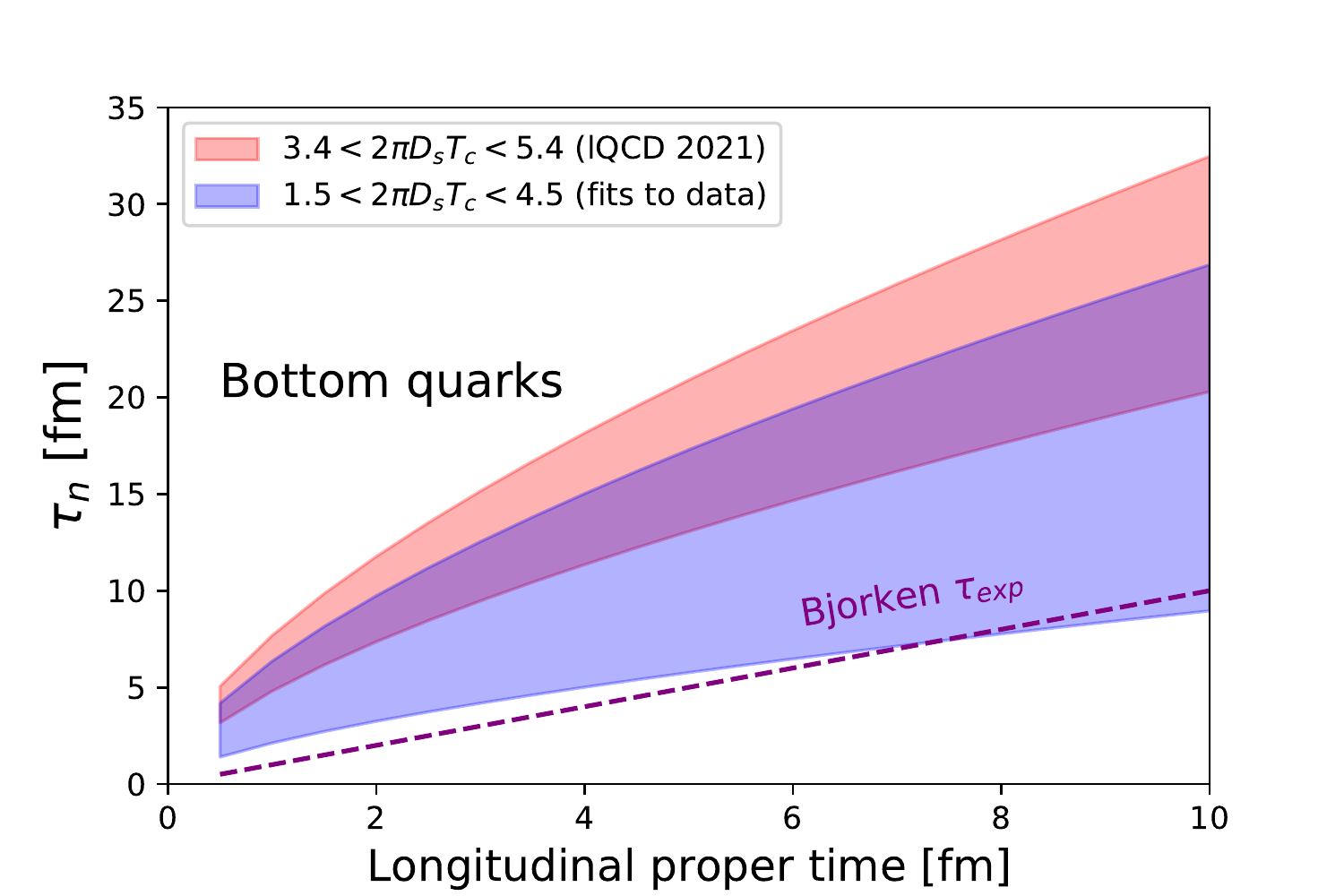}
    \caption{The relaxation time $\tau_n$ of bottom quarks as a function of the longitudinal proper time  is compared to the typical expansion timescale $\tau_{\rm exp}$ of the fluid undergoing a Bjorken flow. 
    }
    \label{fig:bottom_bjorken}
\end{figure}

We now consider the numerical results of our approach.
In Figs.~\ref{fig:charm_bjorken} and~\ref{fig:bottom_bjorken} the comparison between $\tau_{\mathrm{exp}}$ and $\tau_n$ as functions of the longitudinal proper time are reported, respectively for charm and bottom quarks. This is done assuming an initial temperature of 0.45 GeV, initialization time $\tau_0 = 0.5$ fm/c and employing different values of the transport coefficient $D_s$. For charm quarks, we can see that $\tau_n$ goes below $\tau_{\mathrm{exp}}$ quite fast when using transport coefficients arising from fits to experimental data, indicating that the conditions for a fluid-dynamic description are fulfilled for a sizeable fraction of the deconfined fireball lifetime. The Lattice-QCD estimates of the transport coefficients suggest a later hydrodynamization time-scale, but they still allow for a hydrodynamic description after $\sim 5$ fm/c. Regarding bottom quarks, both $D_s$ estimates predict the hydrodynamization time-scale to be of the order of the typical lifetime of the QGP or larger. 

The exact value of $\tau_n$ clearly depends on the initial temperature and formation time of the QGP, as suggested by the estimate in Eq.~(\ref{eq:tauQ_FDR}). The latter are not independent parameters, but are linked by entropy conservation to the final rapidity density of produced particles. One has
\begin{equation}
    T_0^3\tau_0\sim s_0\tau_0\sim\left.\frac{dS_0}{d\eta_s}\right|_{\eta_s=0}\sim\left.\frac{dN}{dy}\right|_{y=0}\;,
\end{equation}
where $S_0$ ($s_0$) is the initial entropy (density), $\eta_s\equiv(1/2)\ln\frac{(t-z)}{(t-z)}$ the spacetime rapidity and $y\equiv(1/2)\ln\frac{(E+p^z)}{(E-p^z)}$ the rapidity of the final detected particles. Hence, according to Eq.~(\ref{eq:tauQ_FDR}), the higher the rapidity density of produced particles, the faster the relaxation of heavy quarks toward equilibrium.

Although the Bjorken flow is not able to describe the full evolution of the plasma but only the first instants after the collision, it still allows us to get a semi-realistic estimate of how fast the diffusion process happens for heavy quarks. Our conclusion is that the applicability of hydrodynamics to the study of charm quark diffusion in the fireball produced in heavy-ion collisions does not seem to be forbidden.

\section{Conclusions and outlook}
\label{sec:conclusions}

Exciting experimental results on charm- and bottom-hadron observables, which have nowadays an unprecedented level of precision, pose the important physics question about the possible heavy-quark thermalization in the QGP~\cite{ALICE:2021rxa}. 
Driven by this question, we adopted a new strategy to study the dynamics of heavy quarks in the QGP and built a connection between a second-order hydrodynamic approach {  based on the heavy-quark current conservation} and the approach provided by transport theory. This led us to an expression for the transport coefficients appearing in the equation of motion of the heavy-quark diffusion current $\tau_n$ and $\kappa_n$ as functions of the temperature of the medium and heavy-quark mass.

Our results display the expected non relativistic limit when $T/M\ll 1$, but can be applied also to heavy quarks with relativistic momenta. Remarkably, within the Fokker-Planck approach, the relation connecting the spatial ($D_s$) and momentum ($D$) diffusion coefficients -- $D_s=T^2/D$ -- holds also in the relativistic domain, as long as the momentum dependence of $D$ can be neglected.

In our approach the conditions for the applicability of hydrodynamics seem to be fulfilled by the charm quark {  for a fraction of the fireball lifetime}, while for the bottom quark the outcome indicates later hydrodynamization.
Our next step will be the implementation of the heavy-quark current in a hydrodynamic framework (FluiduM~\cite{Floerchinger_2014,Floerchinger_2019,Devetak_2020}) to compute and analyse heavy-flavor observables and compare them with experimental data. 
We plan to include the interaction of the heavy-quark current with other conserved currents of baryon number, strangeness and electric charge~\cite{Fotakis_2020,Fotakis:2021vsp} and study how this influences the diffusion process, considering also the presence of strong magnetic fields~\cite{Dubla:2020bdz, ALICE:2019sgg} at the beginning of the collision.

\section*{Acknowledgement}
This work is part of and supported by the DFG Collaborative Research Centre "SFB 1225 (ISOQUANT)".
A.D. is partially supported by the Netherlands Organisation for Scientific Research (NWO) under the grant 19DRDN011, VI.Veni.192.039.

\bibliography{bibliography}

\appendix

\section{Quantum corrections to the Fokker-Planck equation}
\label{app:quantumFP}

In this work we used the Boltzmann and Fokker-Planck equations in their classical limit, namely neglecting quantum corrections associated to the fermionic statistics of heavy quarks (Pauli blocking). Therefore, the distribution function at equilibrium for heavy quarks was expected to be a classical Boltzmann exponential as in Eq.~\eqref{eqn:boltzmanndistribution}. A more accurate estimate for the transport coefficients can be provided by implementing quantum corrections in the Boltzmann equation and in the subsequent Fokker-Planck equation. However, including them can lead to complications concerning the determination of the distribution function of heavy quarks at thermal equilibrium. In fact, finding a stationary solution for the Fokker-Planck equation becomes nontrivial in this case~ \cite{PhysRevE.49.5103}. Nevertheless, if one considers the case of a single momentum-independent diffusion coefficient -- namely $B_0 = B_1 \equiv D$ -- the corresponding Fokker-Planck equation reads
\begin{equation}
\label{eqn:quantum-fokker-planck}
     C[f^{(r)}_{k}] = k^0\frac{\partial}{\partial k^i}\left\{
     A({k})k^if^{(r)}_{k}\Tilde f^{(r)}+D\delta^{ij}\frac{\partial}{\partial k^j}f^{(r)}_{k}]
     \right\}
\,,\end{equation}

\noindent where $\Tilde f^{(r)}_k=1-f^{(r)}_k$ accounts for Pauli blocking. This equation admits an analytical stationary solution in terms of a Fermi-Dirac distribution,

\begin{equation}
f^{(r)}_{0k}= \left[\gamma_Q^{-1}\exp{(\frac{E_k-q_r\mu_Q^{\rm net}}{T})}+1\right]^{-1}\,.
\end{equation}
The relaxation time and diffusion coefficient now read

\begin{gather}
\begin{split}
    \tau_n^{\rm quantum} \sim \frac{I_{31}}{P_0}\frac{T}{D}\left(1+2\frac{1}{3P_0^{(r)}}\int dKk^2 (f_{0k} ^{(r)})^2\right) \\= \tau_n + {\rm correction}\label{eq:quantum1}\,,
\end{split}\\
\begin{split}
  \kappa_n^{\rm quantum} \sim \frac{T^2}{D}n_0^{(r)} \left(1+2\frac{1}{3P_0^{(r)}}\int dK k^2 (f_{0k} ^{(r)})^2\right)\\=\kappa_n + {\rm correction}\label{eq:quantum2} \,.
\end{split}
\end{gather}
Since the correction to the classical value depends on the square of the distribution function, which is exponentially suppressed with $M/T$, we expect a deviation from the classical value only for very small value of $M/T$.

\begin{figure}[ht!]
    \centering
    \includegraphics[width=0.48\textwidth]{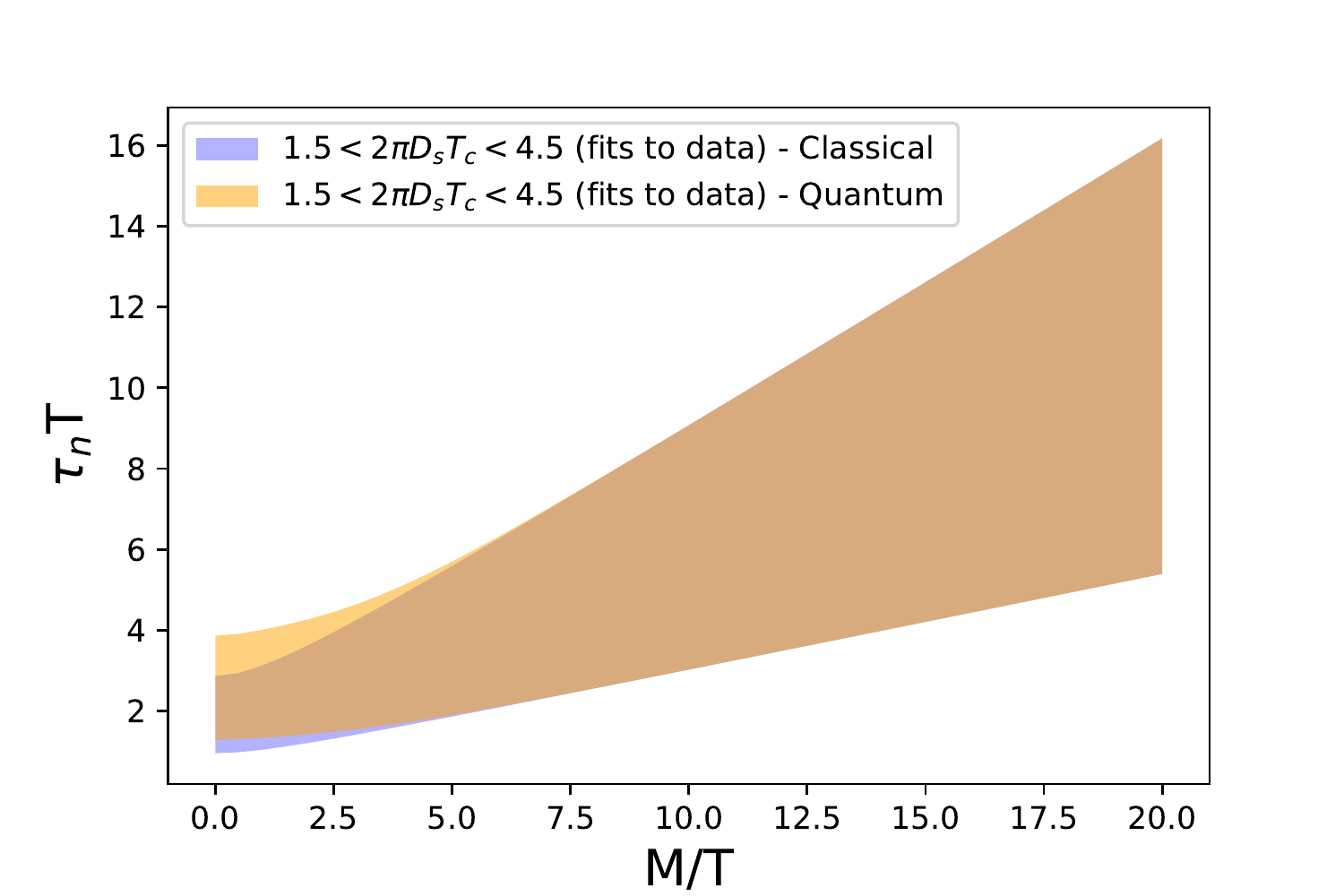}
    \caption{Relaxation time times the temperature as a function of $M/T$. The different bands correspond respectively to the classical and quantum computation of the relaxation time.}
    \label{fig:quantum_tau}
\end{figure}

\begin{figure}[ht!]
    \centering
    \includegraphics[width=0.48\textwidth]{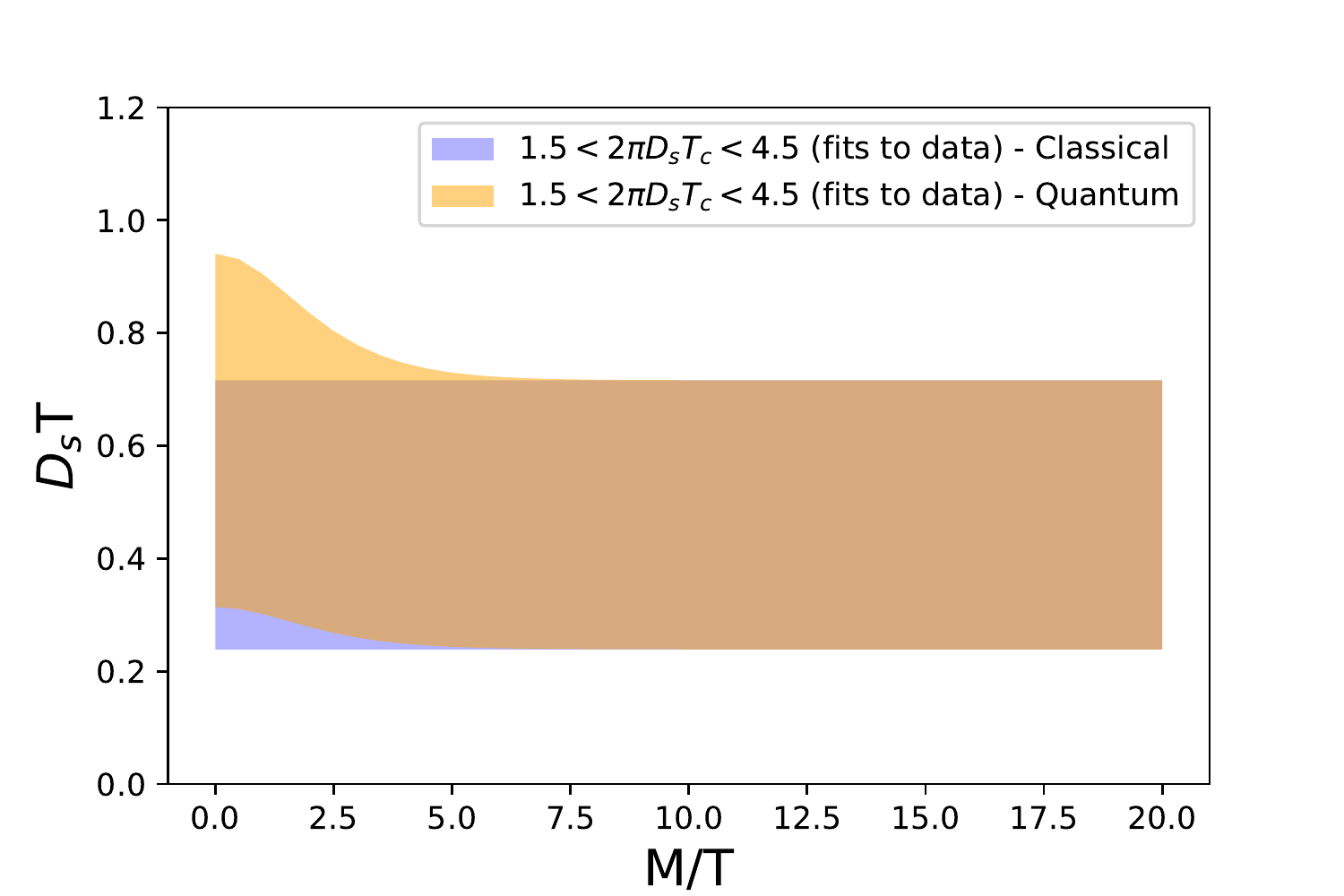}
    \caption{Diffusion coefficient times the temperature as a function of $M/T$. The different bands correspond respectively to the classical and quantum computation of the diffusion coefficient.}
    \label{fig:kappan_quantum}
\end{figure}

Looking at the numerical results for the above coefficients one can see that this is actually the case. 
We start considering the situation of full chemical equilibrium for the heavy quarks, in which $\gamma_Q=1$ and $\mu_Q^{\rm net}=0$.
In Fig.~\ref{fig:quantum_tau} deviations from the classical behaviour in the relaxation time are visible only at very small values of $M/T$. Therefore, they are irrelevant for the realistic conditions realized at the experiment. Similar considerations apply to Fig.~\ref{fig:kappan_quantum}, where only at small $M/T$ values the diffusion coefficient times the temperature differs from its classical constant behavior.

One may worry that for charm quarks at the very beginning of the fireball evolution, when $T\sim 0.5$ GeV, the condition $M/T\gg 1$ is only marginally satisfied. However in this case what matters is that in the early stages charm quarks -- produced in the initial hard scattering processes -- are strongly underpopulated with respect to what would be their equilibrium abundance. This occurrence, discussed in detail in the following section, is quantified by the a fugacity factor $\gamma_Q\ll 1$ which should be included into Eqs.~(\ref{eq:quantum1}) and (\ref{eq:quantum2}). Since the relevance of quantum statistics depends on the $\lambda_{\rm th}/\bar d$ ratio among the thermal de-Broglie wavelength $\lambda_{\rm th}\equiv (2\pi/MT)^{1/2}$ of the particle and the average interparticle distance $\bar d\sim n^{-1/3}$, the classical limit holding when $\lambda_{\rm th}/\bar d\ll 1$, the initial underpopulation of charm quark makes their classical treatment better justified also at the very early stages. The subsequent fireball evolution can only improve the accuracy of the approximation. Considering for simplicity the case of a Bjorken expansion one has
\begin{equation}
    \lambda_{\rm th}\sim T^{-1/2}\sim \tau^{1/6}\quad{\rm and}\quad
    \bar d\sim n^{-1/3}\sim \tau^{1/3}\;,\nonumber
\end{equation}
so that $\lambda_{\rm th}/\bar d\sim \tau^{-1/6}$.

\section{Estimate of the heavy-quark chemical potential in the case of Bjorken flow}
\label{app:fugacityStudy}
In this section we discuss how to fix the heavy-quark chemical potential referring to the conservation of the average heavy-quark number $N_{Q\bar Q}\equiv (N_Q+N_{\bar Q})/2$. The mid-rapidity density at $\tau_0$ arising from the initial hard production is given by
\begin{equation}
n^{Q\bar Q}_{\rm hard}(\tau_0,\vec x_\perp,y=0)=\frac{1}{\tau_0}\left.\frac{d^3N^{Q\bar Q}}{d\vec x_\perp dy}\right|_{y=0}
,\end{equation}
In the above expression, the $Q\bar Q$ rapidity distribution in nucleus-nucleus collisions is set by the pQCD $Q\bar Q$ cross-section
\begin{equation}
    \frac{dN^{Q\bar Q}}{dy}=\langle N_{\rm coll}\rangle \frac{1}{\sigma^{\rm in}}\frac{d\sigma^{Q\bar Q}}{dy}\,,
\end{equation}
where $\sigma^{\rm in}$ is the inelastic proton-proton cross-section and $\sigma^{Q\bar Q}$ is the hard production cross-section, possibly containing cold-nuclear-matter effects (nPDF's). Hence one gets
\begin{equation}
    n^{Q\bar Q}_{\rm hard}(\tau_0,\vec x_\perp,y=0)=\frac{1}{\tau_0} n_{\rm coll}(\vec x_\perp) \frac{1}{\sigma^{\rm in}}\frac{d\sigma^{Q\bar Q}}{dy}\,.
\end{equation}
In case one considers homogeneous conditions in the transverse plane, nevertheless representative of a central Pb-Pb collision, one can estimate:
\begin{equation}
    n^{Q\bar Q}_{\rm hard}(\tau_0,y=0)=\frac{1}{\tau_0} \frac{\langle N_{\rm coll}\rangle}{\pi R_{\rm Pb}^2} \frac{1}{\sigma^{\rm in}}\frac{d\sigma^{Q\bar Q}}{dy}\,.\label{eq:av-dens}
\end{equation}

In order to fix at each point the initial $Q\bar Q$ chemical potential $\mu_{Q}$ ({  the same for quarks and antiquarks, which are produced in equal amount}), this {  density} has to be set equal to the equilibrium thermal multiplicity
\begin{equation} \begin{split}
   n^{Q\bar Q}_{\rm therm}(x)=(2s+1)N_c\left(\frac{MT(x)}{2\pi}\right)^{\frac{3}{2}}\times\\\times e^{-M/T(x)}e^{\mu_{Q}(x)/T(x)}\,.
   \end{split} \end{equation} 
$T(x)$ is extracted from the initial local energy-density of the medium through its Equation of State.
For the sake of simplicity let us introduce the fugacity $\gamma_Q\equiv e^{\mu_{Q}/T}$. One has then:
\begin{equation}
   n^{Q\bar Q}_{\rm therm}(x)=(2s+1)N_c\,\gamma_Q(x)\left(\frac{MT(x)}{2\pi}\right)^{\frac{3}{2}}e^{-M/T(x)}\,.
   \end{equation}
Let us perform some estimates for the initial density of charm-quark pairs with mass $M=1.5$ GeV taking the central prediction by FONLL~\cite{Cacciari:2001td} for collisions at 5.02 TeV. One gets, at $y=0$, $d\sigma^{Q\bar Q}/dy=0.463$ mb, with $\sigma^{\rm in}=70$ mb. For the 0-10\% most central Pb-Pb collisions at $\sqrt{s_{\rm NN}}=5.02$ TeV one has $n_{\rm coll}(\vec x_\perp=0)=31.57\,{\rm fm}^{-2}$ and $\langle N_{\rm coll}\rangle=1653$. Assuming a thermalization time $\tau_0=0.5$ fm/c one gets at the center of the fireball
\begin{equation}
n^{Q\bar Q}_{\rm hard}(\tau_0,\vec x_\perp=0,y=0)\approx 0.42\,{\rm fm}^{-3}\,.    
\end{equation}
The average density in the transverse plane can be estimated as slightly lower. Starting form Eq.~(\ref{eq:av-dens}) and setting $R_{\rm Pb}=6.62$ fm one gets
\begin{equation}
 n^{Q\bar Q}_{\rm hard}(\tau_0,y=0)\approx 0.16\,{\rm fm}^{-3}\,.    
\end{equation}
This has to be compared with the thermal abundance in the case of full chemical equilibrium of the heavy quarks, i.e. $\gamma_Q=1$.
Assuming an initial temperature of the fireball of $T_0=0.45$ GeV one would obtain
\begin{equation}
n^{Q\bar Q}_{\rm chem.eq.}(\tau_0,y=0)\approx 0.98\,{\rm fm}^{-3}\,.    \end{equation}
Initially the heavy quarks are then underpopulated with respect to their chemical-equilibrium abundance. This will be no longer the case at the end of the fireball evolution. The initial heavy-quark fugacity can be estimated as
\begin{equation}
    \gamma_Q(\tau_0)=n^{Q\bar Q}_{\rm hard}(\tau_0)/n^{Q\bar Q}_{\rm chem.eq.}(\tau_0)\approx 0.16\,.
\end{equation}
We now try to estimate the evolution of the heavy-quark density and fugacity while the fireball undergoes an ideal Bjorken expansion. In this case particle conservation entails:
\begin{equation}
    n^{Q\bar Q}(\tau)\tau=n^{Q\bar Q}_0\tau_0\,,
\end{equation}
where $n^{Q\bar Q}_0=n^{Q\bar Q}_{\rm hard}(\tau_0)$. The Landau matching condition applied to the heavy-quark density allows one to extract the heavy-quark fugacity $\gamma_Q(\tau)$:
\begin{equation}
    (2s+1)N_c\,\gamma_Q(\tau)\left(\frac{MT(\tau)}{2\pi}\right)^{\frac{3}{2}}e^{-M/T(\tau)}\frac{\tau}{\tau_0}=n^{Q\bar Q}_0
\end{equation}
In the above, neglecting dissipative effects and deviations from a Stefan-Boltzmann EoS, we estimate the temperature evolution from entropy conservation:
\begin{equation}
    s(\tau)\tau=s_0\tau_0\quad\longrightarrow\quad T^3(\tau)\tau=T_0^3\tau_0
\end{equation}
Let us estimate the value of the heavy-quark fugacity at chemical-freeze-out at $T_{\rm FO}=0.15$ GeV, occurring at $\tau_{\rm FO}=(T_0/T_{\rm FO})^3\tau_0=27\tau_0=13.5$ fm/c. One gets $\gamma_Q(\tau_{\rm FO})\approx 24.6$, not far from the one obtained with SHM fits~\cite{Braun_Munzinger_2000} ($\gamma_c\sim 30$).
\begin{figure}[ht!]
    \centering
    \includegraphics[width=0.48\textwidth]{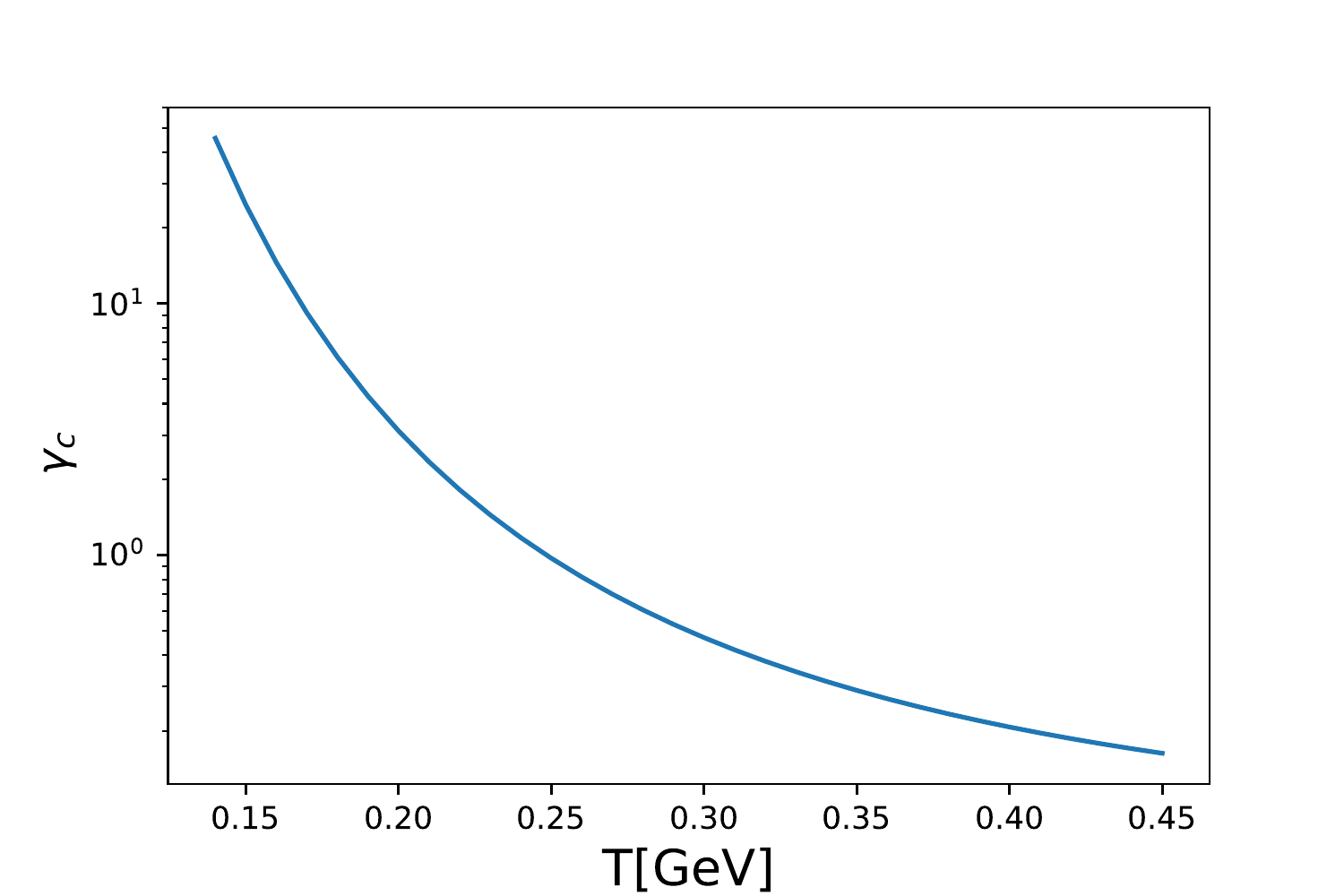}
    \caption{Charm quark fugacity as a function of temperature in logarithmic scale with $T_0=0.45$ GeV.}
    \label{fig:fugValues}
\end{figure}

\section{Coefficients of the linear expansion of the off-equilibrium deviation}
\label{app:calculations}
In this section we determine the coefficients of the linear expansion of the deviation from equilibrium $\delta f_k^{(r)}$ in terms of its moments, expressed in Eq.~\eqref{eq:deviation_exp}. Each coefficient can be computed by integrating the corresponding moment of the deviation $\delta f_k^{(r)}$. 
The orthogonality relations between moments given by~\cite{Denicol_2012},
\begin{equation}
\label{eqapp:orthogonality}
\begin{split}
    \int dK F(k^0)k^{\langle \mu_1...}k^{\mu_n \rangle}k_{\langle \nu_1...}k_{ \nu_m \rangle} =\\= \frac{\delta^m_n m!\Delta^{\mu_1 ... \mu_n}_{\nu_1 ...\nu_m}}{(2m+1)!!}\int {dK F(k^0)(\Delta_{\alpha \beta} k^{\alpha} k^{\beta})^m}\,,   
\end{split}
\end{equation}
are employed. The expansion coefficient for the heavy-quark bulk pressure is obtained from the zeroth moment of the deviation as 

\begin{equation}
\begin{split}
    -\frac{3}{M^2}\Pi_{(r)} = \int dK \delta f_k^{(r)} = \\ =-\frac{3}{M^2} \int dK a_0^{(r)} f_0^{(r)} \Pi_{(r)}\\ \rightarrow a_0^{(r)} = \frac{1}{I_{00}^{(r)}}\,.    
\end{split}
\end{equation}

The coefficient for the heavy-quark diffusion current is computed by taking the first moment of the deviation,

\begin{equation}
\begin{split}
    \nu^{\langle \sigma \rangle}_{(r)} = \int dK k^{\langle \sigma \rangle}  \delta f_k^{(r)} =\\= \int dK f_0^{(r)} a_1^{(r)} k^{\langle \sigma \rangle} k_{\langle \mu \rangle} \nu^\mu_{(r)} =\\ =- \frac{a_1^{(r)}}{3} \delta_\mu^\sigma\nu^\mu_{(r)} \int dK f_0^{(r)} k^2\\ \rightarrow a_1^{(r)} = -\frac{1}{P_0^{(r)}}\,.    
\end{split}
\end{equation}

The coefficient for the heavy-quark shear stress term is obtained by taking the second moment of the deviation,

\begin{equation}
    \begin{split}
    \pi^{\mu \sigma}_{(r)} = \int dK k^{\langle \mu}k^{\sigma \rangle}\delta f_k^{(r)} =\\= \int dK k^{\langle \mu}k^{\sigma \rangle} k_{\langle \alpha}k_{\beta \rangle} a_2^{(r)} f_0^{(r)} \pi_{\alpha \beta}^{(r)} \\ =\frac{2}{15}\int dK  a_2^{(r)} f_0^{(r)} k^4 \pi^{\mu \sigma}_{(r)}\\ \rightarrow a_2^{(r)} = \frac{1}{2 I_{42}^{(r)}} \,, 
\end{split}
\end{equation}

\section{Details on the calculation of the transport coefficients}
\label{app:bulk_corrections}
In this section we report the explicit calculation {  for the heavy-quark} relaxation time and diffusion coefficient {  leading to the result} in Eq.~\eqref{eqn:taun}.
The starting point is the Fokker-Planck equation for the heavy (anti)quark distributions (charm, anti-charm, bottom, anti-bottom)
\begin{equation}
\label{eqapp:fp}
    k^\mu \partial_\mu f_k^{(r)} = k_0\frac{\partial}{\partial k^i}\left\{
     {A k^i}f^{(r)}_{k}+\delta^{ij}D\frac{\partial}{\partial k^j}f^{(r)}_{k}
     \right\}\,,
\end{equation}
where we consider {  the case of a isotropic momentum broadening, i.e.} $D\!=\!B_0\!=\!B_1$.

The zeroth moment of the Fokker-Planck equation gives the continuity equation in the LRF of the fluid,
\begin{equation}
    \partial_t n_{(r)} +\partial_i \nu^i_{(r)} = 0 \quad\longrightarrow\quad \partial_t n_+ + \partial_i \nu^i_+ = 0\,.
\end{equation}
Notice that the RHS of Eq.~\eqref{eqapp:fp} {  provides a vanishing contribution} when taking its zeroth moment. This can be verified by doing the integration by parts.

The first moment of the Fokker-Planck equation gives
\begin{equation}
\begin{split}
  \partial_t \int dK k_0 k^l f_k^{(r)}   +\partial_i\int dK {k^l k^i} f_k^{(r)} \\
 =\int dK k_0 k^l \frac{\partial}{\partial k^i}\left ( A k^i f_k^{(r)} \right)\label{eq:FP1st}  
\end{split}
\end{equation}
As we will show below, this will lead to the equation of motion for the diffusion current in the LRF of the fluid.
Notice that the term proportional to the momentum-broadening coefficient vanishes when {  taking} the first moment {of the Fokker-Planck equation}. In fact, since it is proportional to a second-order derivative it vanishes {  after integration} by parts. Let us now analyze all the terms {  involved in Eq. (\ref{eq:FP1st})} separately.

\subsection{1st term}
Here we compute the term containing the time derivative of the distribution function,
\begin{equation}
    \partial_t \int dK k_0 k^l f_k^{(r)}\,.
\end{equation}
Due to symmetry properties of the distribution function at equilibrium (it depends only on the particle energy in the LRF of the fluid), the first moment of $f_{0k}^{(r)}$ vanishes. The only contribution comes from the {  off-equilibrium} deviation $\delta f_k^{(r)}$, which we expand in terms of the diffusion current,
\begin{equation}
\partial_t \int dK k_0 k^l f_0^{(r)}\left(-\frac{1}{P_0^{(r)}}k_{\langle \mu\rangle}\nu^{\langle \mu \rangle}\right)\,.
\end{equation}
We then employ the orthogonality relation in Eq.~\eqref{eqapp:orthogonality}\,
\begin{equation}
\partial_t \int dK \frac{k^2}{3} k_0 f_0^{(r)}\left(\frac{1}{P_0^{(r)}}\right)\nu^l\,, 
\end{equation}
and, rewriting in terms of the thermodynamic integrals {introduced in the text}, we get
\begin{equation}
\frac{I_{31}^{(r)}}{P_0^{(r)}}\partial_t \nu^l_{(r)} \quad\longrightarrow\quad\frac{I_{31}}{P_0}\partial_t \nu^l_+\,,
\end{equation}
where 
\begin{equation}
I_{31}^{(r)} = \frac{1}{3}\langle k_0 k^2 \rangle_{0,r}\,.  
\end{equation}
 Notice that $I_{31} \sim M P_0$ in the non-relativistic limit, reducing the computed term to $M\partial_t \nu^l_+$. 

\subsection{2nd term}
Here we compute the term containing the spatial derivative of the distribution function,
\begin{equation}
  \partial_i\int dK {k^l k^i} f_k^{(r)}\,. 
\end{equation}
We use the decomposition for the distribution function to get
\begin{equation}
\partial_i\,\delta^{il}\!\!\int dK \frac{k^2}{3} f_0^{(r)} +
\partial_i \int dK {k^{i} k^{l}}\delta f_k^{(r)}\;.  
\end{equation}
Exploiting the orthogonality conditions and the definition of the pressure, we get
\begin{equation}
\begin{split}
\delta^{il}\partial_i  P_0^{(r)} + O(\delta^{il}\partial_i \Pi) + O(\partial_i \pi^{il}) \\ 
    =T   n_0^{(r)}\delta^{il}\partial_i \left(\frac{\mu_r}{T}\right)+ {\rm \, corr}    
\end{split}
\end{equation}
where in the last passage we used $\partial_i P_0 = T n_0 \partial_i (\mu_r/T)$ {  and the neglected terms, involving derivatives of the bulk pressure and of the shear stress, are at least of second order in the gradients}.

\subsection{3rd term}
Here we compute the RHS of the equation. Notice that the term containing the momentum-diffusion coefficient doesn't contribute. In fact, it is proportional to a second order derivative, thus its first moment vanishes. Hence, one has simply to compute
\begin{equation}
    \begin{split}
    \int dK k^l k_0 \frac{\partial}{\partial k^i}\left ( A k^i f^{(r)} \right) \\
    =\int \frac{d^3k}{(2\pi)^3} k^l \left[\frac{\partial k^i}{\partial k^i}\left ( A f^{(r)} \right)+  \frac{\partial A f^{(r)}}{\partial k^i}k^i\right]\\
    \stackrel{\rm IBP}{=}\int \frac{d^3k}{(2\pi)^3}  \left[3\left ( A f^{(r)} \right) k^l -  \frac{\partial (k^i k^l)}{\partial k^i}A f^{(r)}\right] \\
    = \int \frac{d^3k}{(2\pi)^3} \Big[3\left ( A f^{(r)} \right)k^l -  3\left(A f^{(r)}\right)k^l +\\  
    - k^l A f^{(r)}\Big] 
    = \int \frac{d^3k}{(2\pi)^3} \left[ - k^l A f^{(r)}\right]
    \end{split}
\end{equation}
where $\rm{IBP}$ means we performed the integration by parts. Now we exploit the decomposition of the distribution function. Due to symmetry constraints, the equilibrium part of the distribution doesn't contribute since its first moment is zero. Thus we have
\begin{equation}
    \int dK k^0 k^l A f_0^{(r)} \frac{k_{\langle \mu\rangle}\nu^{\mu}}{P_0^{(r)}}\,.
\end{equation}
By exploiting the orthogonality relation, one obtains
\begin{equation}
    \begin{split}
    -\frac{1}{3P_0^{(r)}}\int dK k^0 k^2 A f_0^{(r)}\nu^l   \\ =
    -  \frac{1}{P_0^{(r)}}\frac{1}{3}\int dK k^0 k^2 \left( \frac{D}{k^0 T}\right) f_0^{(r)}\nu^l  \\
    =\frac{D}{\cancel{P_0^{(r)}} T}\cancel{\left[\frac{1}{3}\int dK k^2 f_0^{(r)}\right]}\nu^l  \\
    =- \frac{D}{T} \nu^l_{(r)}\,,
    \end{split}
\end{equation}
where we made use of the Einstein fluctuation-dissipation relation to express $A$ in terms of the momentum-diffusion coefficient $D$.

\subsection{Putting all blocks together}
We now combine the three terms to obtain the equation for the diffusion current:
\begin{equation}
  \frac{T}{D} \frac{I_{31}}{P_0} \partial_t \nu^l_{(r)} + \nu^l_{(r)} = -\frac{T^2}{D}n_0^{(r)} \partial_l \left(\frac{\mu_r}{T}\right)\,.
\end{equation}
This is a relaxation-type equation for the diffusion current $\nu_{(r)}^\mu$.
Thus, we can identify the corresponding relaxation time and diffusion coefficient,
\begin{gather}
    \tau_n = \frac{T I_{31}}{D P_0}\,,\\
    \kappa_n^{(r)} = \frac{T^2}{D}n_0^{(r)} \equiv D_s n_0^{(r)}\,.
\end{gather}
We find that the relation $D_s = T^2/D$ between the spatial ($D_s$) and momentum ($D$) diffusion coefficients, usually found in studying the non-relativistic Brownian motion, 
arises naturally and holds also in this case in which the heavy particle undergoes a relativistic dynamics, with $E_k=\sqrt{k^2+M^2}$. This is a non-trivial result, valid as long as the momentum dependence of $D$ can be neglected.

\end{document}